\documentclass[a4paper,10pt]{revtex4}
\usepackage{graphicx}  
\usepackage{amsmath}   
\usepackage{amssymb}   
\usepackage{bm} 
\usepackage{dcolumn}
\usepackage{color}
\usepackage{mathrsfs}
\usepackage{amsfonts}
\usepackage{varioref}
\usepackage{mathrsfs}
\usepackage{graphicx}
\usepackage{latexsym}
\usepackage{amsmath}
\usepackage{amssymb}
\usepackage{textcomp}
\usepackage{amsbsy}
\usepackage{graphics}
\usepackage{epstopdf}
\usepackage{color}
\usepackage[caption=false]{subfig}

\RequirePackage[colorlinks,citecolor=blue,urlcolor=magenta,linkcolor=blue]{hyperref}
\input epsf

\allowdisplaybreaks[4]

\begin{document}

\tolerance=5000

\title{Generalised (non-singular) entropy functions with applications to cosmology and black holes}

\author{Sergei~D.~Odintsov$^{1,2}$\,\thanks{odintsov@ieec.uab.es},
Tanmoy~Paul$^{3}$\,\thanks{pul.tnmy9@gmail.com}} \affiliation{
$^{1)}$ ICREA, Passeig Luis Companys, 23, 08010 Barcelona, Spain\\
$^{2)}$ Institute of Space Sciences (ICE, CSIC) C. Can Magrans s/n, 08193 Barcelona, Spain\\
$^{3)}$ Department of Physics, Chandernagore College, Hooghly - 712 136, India.}


\tolerance=5000

\begin{abstract}
The growing interest of different entropy functions proposed so far 
(like the Bekenstein-Hawking, Tsallis, R\'{e}nyi, Barrow, Sharma-Mittal, Kaniadakis and Loop Quantum Gravity entropies) towards black hole thermodynamics 
as well as towards cosmology lead to the natural question that whether there exists a generalized entropy function that can generalize 
all these known entropies. With this spirit, 
we propose a new 4-parameter entropy function that seems to converge to the aforementioned known entropies for certain limits of the entropic parameters. 
The proposal of generalized entropy is extended to non-singular case, in which case , the entropy proves to be singular-free 
during the entire cosmological evolution of the universe. The hallmark of such generalized entropies is that it helps us to fundamentally understand 
one of the important physical quantities namely ``entropy''. Consequently we address the implications of the generalized entropies on black hole thermodynamics as well as on cosmology, 
and discuss various constraints of the entropic parameters from different perspectives. 
\end{abstract}

\maketitle

\section{Introduction}

One of the most important discoveries in theoretical physics is the black body radiation of a black hole, which is described by a certain temperature and 
by a Bekenstein-Hawking entropy function \cite{Bekenstein:1973ur,Hawking:1975vcx} (see 
\cite{Bardeen:1973gs,Wald:1999vt} for extensive reviews). 
On contrary to classical thermodynamics where the entropy is proportional to volume of the system 
under consideration, the Bekenstein-Hawking entropy is proportional to the area of the black hole horizon. Such unusual behaviour of the black hole 
entropy leads to the proposals of different entropy functions, such as, the Tsallis \cite{Tsallis:1987eu}, R\'{e}nyi \cite{Renyi}, 
Barrow \cite{Barrow:2020tzx}, 
Sharma-Mittal \cite{SayahianJahromi:2018irq}, Kaniadakis \cite{Kaniadakis:2005zk} and the Loop Quantum 
Gravity entropies \cite{Majhi:2017zao} are well known entropy functions proposed so far. All of these known entropies have the common properties like -- 
(1) they seem to be the monotonic increasing function with respect to the Bekenstein-Hawkinh entropy variable, (2) they obey the third law of 
thermodynamics, in particular, all of these entropies tend to zero as $S \rightarrow 0$ (where $S$ represents the Bekenstein-Hawking entropy) and 
(3) they converge to the Bekenstein-Hawking entropy for suitable choices of the respective entropic parameter, for example, the Tsallis entropy 
goes to the Bekenstein-Hawking entropy when the Tsallis exponent tends to unity. Furthermore, these entropies have rich consequences 
towards cosmology, particularly in describing the dark energy era of the universe 
\cite{Li:2004rb,Li:2011sd,Wang:2016och,Pavon:2005yx,Nojiri:2005pu,Landim:2022jgr,Zhang:2005yz,Guberina:2005fb,Elizalde:2005ju,
Ito:2004qi,Gong:2004cb,BouhmadiLopez:2011xi,Malekjani:2012bw,Khurshudyan:2016gmb,Landim:2015hqa,
Gao:2007ep,Li:2008zq,Anagnostopoulos:2020ctz,Zhang:2005hs,Li:2009bn,Feng:2007wn,Zhang:2009un,Lu:2009iv,
Micheletti:2009jy,Mukherjee:2017oom,Nojiri:2017opc,Nojiri:2019skr,
Saridakis:2020zol,Barrow:2020kug,Adhikary:2021xym,Srivastava:2020cyk,Bhardwaj:2021chg,Chakraborty:2020jsq,Sarkar:2021izd}. The growing interest of such known entropies 
and due to their common properties lead to a natural question that whether there exists some generalized entropy function which is able to generalize 
all the known entropies proposed so far for suitable limits of the parameters. 

The entropy functions are extensively applied in the realm of black hole thermodynamics and cosmological evolution of the universe. Recently 
we showed that the entropic cosmology corresponding to different entropy functions can be equivalently represented by holographic cosmology 
where the equivalent holographic cut-offs come in terms of either particle horizon and its derivative or the future horizon and its derivative. One 
of the mysteries in today's cosmology is to explain the acceleration of the universe in the high as well as in the low curvature regime, 
known as inflation and the dark energy era respectively. These eras are well described by entropic cosmology or equivalently by 
holographic cosmology 
\cite{Li:2004rb,Li:2011sd,Wang:2016och,Pavon:2005yx,Nojiri:2005pu,Landim:2022jgr,Zhang:2005yz,Guberina:2005fb,Elizalde:2005ju,
Ito:2004qi,Gong:2004cb,BouhmadiLopez:2011xi,Malekjani:2012bw,Khurshudyan:2016gmb,Landim:2015hqa,
Gao:2007ep,Li:2008zq,Anagnostopoulos:2020ctz,Zhang:2005hs,Li:2009bn,Feng:2007wn,Zhang:2009un,Lu:2009iv,
Micheletti:2009jy,Mukherjee:2017oom,Nojiri:2017opc,Nojiri:2019skr,
Saridakis:2020zol,Barrow:2020kug,Adhikary:2021xym,Srivastava:2020cyk,Bhardwaj:2021chg,Chakraborty:2020jsq,Sarkar:2021izd,
Nojiri:2022aof,Nojiri:2022dkr,Horvat:2011wr,Nojiri:2019kkp,Paul:2019hys,Bargach:2019pst,Elizalde:2019jmh,Oliveros:2019rnq,Mohammadi:2022vru,
Chakraborty:2020tge}, 
and more interestingly, the entropic cosmology proves to be useful to unify the early inflation 
and the late dark energy era of the universe in a covariant manner \cite{Nojiri:2020wmh}. 
Apart from the inflation, the holographic cosmology turns out to be useful 
in describing the bouncing scenario \cite{Nojiri:2019yzg,Brevik:2019mah}. 
In regard to the bounce scenario, the energy density sourced from the holographic principle or from 
some entropy function under consideration helps to violate the null energy condition at a finite time, which in turn triggers a 
non-singular bouncing universe. However here it deserves mentioning that all the known entropies mentioned above (like 
Tsallis , R\'{e}nyi, Barrow, Sharma-Mittal, Kaniadakis and the Loop Quantum Gravity entropies) become singular (or diverge) at a certain cosmological 
evolution of the universe, particularly in the context of bounce cosmology. Actually such entropies contain a factor that is proportional to 
$1/H^2$ (where $H$ is the Hubble parameter), and thus they diverge at the instant when the Hubble parameter vanishes, i.e, at the instant of a bounce 
in bouncing cosmology. This makes such known entropies ill-defined in describing a non-singular bounce scenario. 

Based on the above arguments, the questions that naturally arise are following:
\begin{itemize}
 \item Does there exist a generalized entropy function that generalizes all the known entropies proposed so far ?
 \item If so, then what is its implications on black hole thermodynamics as well as on cosmology ?
 \item Similar to the known entropies, is the generalized entropy becomes singular at the instant 
 when the Hubble parameter of the universe vanishes, for instance, in the bounce cosmology ? If so, then does there exist 
 an entropy function that generalizes all the known entropies, and at the same time, also proves to be singular-free during the entire cosmic 
 evolution of the universe ?
\end{itemize}

The present article, based on some of our previous works \cite{Nojiri:2022aof,Nojiri:2022dkr,OP-submitted},
gives a brief review in answering the above questions. The notations or conventions 
in this article are following: we will follow the $\left(-,+,+,+\right)$ signature of the spacetime metric, and 
$\kappa^2 = 8\pi G = \frac{1}{M_\mathrm{Pl}^2}$ where $G$ is the Newton's constant or $M_\mathrm{Pl}$ denotes the four dimensional 
Planck mass. 
In regard to the cosmological evolution, 
$a(t)$ and $H(t)$ are the scale factor and the Hubble parameter of the universe respectively, $N$ being the e-folding number, 
an overprime will denote $\frac{d}{d\eta}$ where 
$\eta$ is the conformal time, an overdot will symbolize $\frac{d}{dt}$ with $t$ being the cosmic time, otherwise an overprime with some argument 
will represent the derivative of the function with respect to that argument.

\section{Possible generalizations of known entropies}
\label{sec:2}
Here we will propose a generalized four-parameter entropy function which can lead to various known entropy functions proposed so far 
for suitable choices of the parameters.

Let us start with the Bekenstein-Hawking entropy, the very first proposal of thermodynamical entropy of 
black hole physics \cite{Bekenstein:1973ur,Hawking:1975vcx},
\begin{align}
S = \frac{A}{4G} \,,
\label{BH-entropy}
\end{align}
where $A = 4\pi r_h^2$ is the area of the horizon and $r_h$ is the horizon radius. 
Consequently, different entropy functions have been introduced 
depending on the system under consideration. Let us briefly recall some of the entropy functions proposed so far:

\begin{itemize}
\item For the systems with long range interactions where the Boltzmann-Gibbs entropy is not applied, one needs to introduce 
the Tsallis entropy which is given by \cite{Tsallis:1987eu},
\begin{align}
S_T = \frac{A_0}{4G}\left(\frac{A}{A_0}\right)^{\delta} \,,
\label{Tsallis entropy}
\end{align}
where $A_0$ is a constant and $\delta$ is the exponent.
\item The R\'{e}nyi entropy is given by \cite{Renyi},
\begin{align}
S_\mathrm{R} = \frac{1}{\alpha} \ln \left( 1 + \alpha S \right) \,,
\label{Renyi entropy}
\end{align}
where $S$ is identified with the Bekenstein-Hawking entropy and $\alpha$ is a parameter.  
\item The Barrow entropy is given by \cite{Barrow:2020tzx},
\begin{align}
\label{Barrow-entropy}
S_\mathrm{B} = \left(\frac{A}{A_\mathrm{Pl}} \right)^{1+\Delta/2} \,,
\end{align}
where $A$ is the usual black hole horizon area and $A_\mathrm{Pl} = 4G$ is the Planck area. 
The Barrow entropy describes the fractal structures of black hole that may generate from quantum gravity effects.  
\item The Sharma-Mittal entropy is given by \cite{SayahianJahromi:2018irq},
\begin{align}
S_{SM} = \frac{1}{R}\left[\left(1 + \delta ~S\right)^{R/\delta} - 1\right] \,,
\label{SM entropy}
\end{align}
where $R$ and $\delta$ are two parameters. 
The Sharma-Mittal entropy can be regarded as a possible combination of the Tsallis and R\'{e}nyi entropies. 
\item The Kaniadakis entropy function is of the following form \cite{Kaniadakis:2005zk}:
\begin{align}
S_K = \frac{1}{K}\sinh{\left(KS\right)} \,,
\label{K-entropy}
\end{align}
where $K$ is a phenomenological parameter.
\item In the context of Loop Quantum Gravity, one may get the following entropy function 
\cite{Majhi:2017zao}:
\begin{align}
S_q = \frac{1}{\left(1-q\right)}\left[\mathrm{e}^{(1-q)\Lambda(\gamma_0)S} - 1\right] \,,\label{LQG entropy}
\end{align}
where $q$ is the exponent and $\Lambda(\gamma_0) = \ln{2}/\left(\sqrt{3}\pi\gamma_0\right)$ with $\gamma_0$ being the Barbero-Immirzi parameter. 
The $\gamma_0$ generally takes either $\gamma_0 = \frac{\ln{2}}{\pi\sqrt{3}}$ or $\gamma_0 = \frac{\ln{3}}{2\pi\sqrt{2}}$. 
However with $\gamma_0 = \frac{\ln{2}}{\pi\sqrt{3}}$, $\Lambda(\gamma_0)$ becomes unity 
and $S_q$ resembles with the Bekenstein-Hawking entropy for $q \rightarrow 1$. 
\end{itemize}
All the above entropies -- (1) obeys the generalized third law of thermodynamics, i.e the entropy function(s) vanishes at the limit 
$S \rightarrow 0$; (2) monotonically increases with respect to the Bekenstein-Hawking variable and (3) converges to the Bekenstein-Hawking entropy 
for suitable limit of the entropic parameter, for example, the Tsallis entropy tends to $S$ at $\delta = 1$. 

In \cite{Nojiri:2022aof,Nojiri:2022dkr}, we proposed two different entropy functions containing 6-parameters and 4-parameters respectively, 
which can generalize all the known entropies mentioned from Eq.(\ref{Tsallis entropy}) to Eq.(\ref{LQG entropy}). In particular, 
the generalized entropies are given by,
\begin{eqnarray}
\mathrm{6~parameter~entropy:}~~~~~\mathcal{S}_\mathrm{6} \left[ \alpha_\pm, \beta_\pm, \gamma_\pm \right]&=&\frac{1}{\alpha_+ + \alpha_-}
\left[ \left( 1 + \frac{\alpha_+}{\beta_+} \, \mathcal{S}^{\gamma_+}
\right)^{\beta_+} - \left( 1 + \frac{\alpha_-}{\beta_-}
\, \mathcal{S}^{\gamma_-} \right)^{-\beta_-} \right] \,,\label{6-entropy}\\
\mathrm{4~parameter~entropy:}~~~~~S_\mathrm{g}\left[\alpha_+,\alpha_-,\beta,\gamma \right]&=&\frac{1}{\gamma}\left[\left(1 + \frac{\alpha_+}{\beta}~S\right)^{\beta} 
 - \left(1 + \frac{\alpha_-}{\beta}~S\right)^{-\beta}\right] \,,
\label{gen-entropy}
\end{eqnarray}
where the respective parameters are given in the argument and they are assumed to be positive. Here $S$ is the Bekenstein-Hawking entropy. 
Below we prove the generality of the above generalized entropy functions, in particular, 
we show that both the generalized entropies reduce to the known entropies mentioned in Eqs.~(\ref{Tsallis entropy}), 
(\ref{Renyi entropy}), (\ref{Barrow-entropy}), (\ref{SM entropy}), (\ref{K-entropy}), and (\ref{LQG entropy}) for suitable choices of the respective 
parameters. Here we establish it particularly for the 4-parameter entropy function, while the similar calculations hold for the 6-parameter entropy as well 
\cite{Nojiri:2022aof}.

\begin{itemize}
\item For $\alpha_+ \rightarrow \infty$ and $\alpha_- = 0$, one gets
\begin{align}
S_\mathrm{g} = \frac{1}{\gamma}\left(\frac{\alpha_+}{\beta}\right)^{\beta}S^{\beta} \,.\nonumber
\end{align}
If we further choose $\gamma = \left(\alpha_+/\beta\right)^{\beta}$, then the generalized entropy reduces to
\begin{align}
 S_\mathrm{g} = S^{\beta} \,.\nonumber
\end{align}
Therefore with $\beta = \delta$ or $\beta = 1 + \Delta$, the generalized entropy resembles with the Tsallis entropy or with the Barrow entropy 
respectively.

\item For $\alpha_- = 0$, $\beta \rightarrow 0$ and $\frac{\alpha_+}{\beta} \rightarrow \mathrm{finite}$ -- Eq.~(\ref{gen-entropy}) leads to,
\begin{align}
 S_\mathrm{g} = \frac{1}{\gamma}\left[\left(1 + \frac{\alpha_+}{\beta}~S\right)^{\beta} - 1\right] 
 = \frac{1}{\gamma}\left[\exp{\left\{\beta\ln{\left(1 + \frac{\alpha_+}{\beta}~S\right)}\right\}} - 1\right] 
 \approx \frac{1}{\left(\gamma/\beta\right)} \ln{\left(1 + \frac{\alpha_+}{\beta}~S\right)} \,.\nonumber
\end{align}
Further choosing $\gamma = \alpha_+$ and identifying $\frac{\alpha_+}{\beta} = \alpha$, we can write the above expression as,
\begin{align}
 S_\mathrm{g} = \frac{1}{\alpha}\ln{\left(1 + \alpha~S\right)} \,,
\end{align}
i.e.,  $S_\mathrm{g}$ reduces to the R\'{e}nyi entropy.

\item In the case when $\alpha_- = 0$, the generalized entropy becomes,
\begin{align}
 S_\mathrm{g} = \frac{1}{\gamma}\left[\left(1 + \frac{\alpha_+}{\beta}~S\right)^{\beta} - 1\right] \,.
\end{align}
Thereby identifying $\gamma = R$, $\alpha_+ = R$ and $\beta = R/\delta$, the generalized entropy function $S_\mathrm{g}$ gets similar to the 
Sharma-Mittal entropy.

\item For $\beta \rightarrow \infty$, $\alpha_+ = \alpha_- = \frac{\gamma}{2} = K$, we may write Eq.~(\ref{gen-entropy}) as,
\begin{align}
S_\mathrm{g}=&\, \frac{1}{2K}\lim_{\beta \rightarrow \infty}\left[\left(1 + \frac{K}{\beta}~S\right)^{\beta} 
 - \left(1 + \frac{K}{\beta}~S\right)^{-\beta}\right]\nonumber\\
=&\, \frac{1}{2K}\left[\mathrm{e}^{KS} - \mathrm{e}^{-KS}\right] = \frac{1}{K}\sinh{\left(KS\right)} 
\rightarrow \mathrm{Kaniadakis~entropy} \,.
\end{align}

\item Finally, with $\alpha_- = 0$, $\beta \rightarrow \infty$ and $\gamma = \alpha_+ = (1-q)$, Eq.~(\ref{gen-entropy}) immediately yields,
\begin{align}
 S_\mathrm{g} = \frac{1}{(1-q)}\left[\mathrm{e}^{(1-q)S} - 1\right] \,,\nonumber
\end{align}
which is the Loop Quantum Gravity entropy with $\Lambda(\gamma_0) = 1$ or equivalently $\gamma_0 = \frac{\ln{2}}{\pi\sqrt{3}}$.

\end{itemize}
Furthermore, the generalized entropy function in Eq.~(\ref{gen-entropy}) shares the following properties: 
(1) $S_\mathrm{g} \rightarrow 0$ for $S \rightarrow 0$. (2) The entropy 
$S_\mathrm{g}\left[ \alpha_+,\alpha_-,\beta,\gamma \right]$ is a monotonically increasing function with $S$ because both the terms 
$\left(1 + \frac{\alpha_+}{\beta}~S\right)^{\beta}$ and $-\left(1 + \frac{\alpha_-}{\beta}~S\right)^{-\beta}$ present in the expression 
of $S_\mathrm{g}$ 
increase with $S$. (3) $S_\mathrm{g}\left[ \alpha_+,\alpha_-,\beta,\gamma \right]$ seems to converge to the Bekenstein-Hawking 
entropy at certain limit of the parameters. 
In particular, for $\alpha_+ \rightarrow \infty$, $\alpha_- = 0$, $\gamma = \left(\alpha_+/\beta\right)^{\beta}$ and $\beta = 1$, 
the generalized entropy function 
in Eq.~(\ref{gen-entropy}) becomes equivalent to the Bekenstein-Hawking entropy.

Here it deserves mentioning that beside the entropy function proposed in Eq.~(\ref{gen-entropy}) which contains four parameters, one may consider 
a three parameter entropy having the following form:
\begin{align}
S_3[\alpha,\beta,\gamma] = \frac{1}{\gamma}\left[\left(1 + \frac{\alpha}{\beta}~S\right)^{\beta} - 1\right] \,,
\label{3-parameter-entropy}
\end{align}
where $\alpha$, $\beta$ and $\gamma$ are the parameters. 
The above form of $S_3[\alpha,\beta,\gamma]$ satisfies all the properties, 
like -- (1) $S_3[\alpha,\beta,\gamma] \rightarrow 0$ for $S \rightarrow 0$, (2) $S_3$ is an increasing function with $S$ and (3) $S_3$ 
has a Bekenstein-Hawking entropy limit for the choices: $\alpha \rightarrow \infty$, $\gamma = \left(\alpha/\beta\right)^{\beta}$ and $\beta = 1$ 
respectively. 
However $S_3[\alpha,\beta,\gamma]$ is not able to generalize all the known entropies mentioned from Eq.~(\ref{Tsallis entropy}) 
to Eq.~(\ref{LQG entropy}), in particular, $S_3[\alpha,\beta,\gamma]$ does not reduce to the Kaniadakis entropy for any possible 
choices of the parameters.\\

\textbf{\underline{Conjecture - I}}: 
Based on our findings, we propose the following postulate in regard to the generalized entropy function -- 
``The minimum number of parameters required in a generalized entropy function that can generalize 
all the known entropies mentioned from Eq.~(\ref{Tsallis entropy}) to Eq.~(\ref{LQG entropy}) is equal to four''.\\

Below we will address the possible implications of such generalized entropies on black hole thermodynamics as well as on cosmology.

\section{Black hole thermodynamics with 3-parameter generalized entropy}
\label{sec:3}

It is interesting to see what happens when the generalized
entropy~(\ref{3-parameter-entropy}) is ascribed to the prototypical black hole, given
by the Schwarzschild geometry \cite{Nojiri:2022aof}
\begin{align}
\label{dS3BB}
ds^2 = - f(r) \, dt^2 + \frac{dr^2}{f(r)} + r^2 d\Omega^2_{(2)}\, ,
\quad\quad
f(r) = 1 - \frac{2GM}{r} \,,
\end{align}
where  $M$ is the black hole mass and $d\Omega^2_{(2)}=d\vartheta^2
+\sin^2 \vartheta \, d\varphi^2$ is the line element on  the unit
two-sphere.  The black hole event horizon is located at the Schwarzschild
radius
\begin{align}
\label{horizonradius}
r_\mathrm{H}=2GM\, .
\end{align}
Studying quantum field theory on the spacetime with this horizon,
Hawking discovered that the Schwarzschild black hole radiates with a
blackbody spectrum at the temperature 
\begin{align}
\label{dS6BB}
T_\mathrm{H} = \frac{1}{8\pi GM}\,.
\end{align}
  As explained in general below, if we assume that the mass
$M$ coincides with the thermodynamical energy, then the temperature
obtained from the thermodynamical law is different from the Hawking
temperature, a contradiction for observers detecting Hawking radiation.
Alternatively, if the Hawking temperature $T_\mathrm{H}$ is identified
with the physical black hole temperature, the obtained thermodynamical
energy differs from the Schwarzschild mass $M$ even for the Tsallis
entropy or the R{\'e}nyi entropy, which seems to imply a breakdown of
energy conservation.

If the mass $M$ coincides with the thermodynamical energy $E$ of the
system due to energy conservation, as in,
in order for this system to be consistent with the thermodynamical
equation $d\mathcal{S}_G=dE/T$ one needs to define the
generalized temperature $T_\mathrm{G}$ as
\begin{align}
\label{TR1}
\frac{1}{T_\mathrm{G}} \equiv \frac{d\mathcal{S}_\mathrm{G}}{dM}
\end{align}
which is, in general, different from the Hawking temperature $T_\mathrm{H}$.
For example, in the case of the entropy~(\ref{3-parameter-entropy}), one
has
\begin{align}
\label{TR1B}
\frac{1}{T_\mathrm{G}}
= \frac{\alpha}{\gamma} \left( 1 + \frac{\alpha}{\beta} \, \mathcal{S}
\right)^{\beta-1} \frac{d\mathcal{S}}{dM}
= \frac{\alpha}{\gamma}  \left( 1 + \frac{\alpha}{\beta} \, \mathcal{S}
\right)^{\beta-1} \frac{1}{T_\mathrm{H}} \,,
\end{align}
where
\begin{align}
\label{SMT}
\mathcal{S}=\frac{A}{4G}= 4 \pi G M^2 = \frac{1}{16\pi G{T_\mathrm{H}}^2}
\,.
\end{align}
Because $\frac{\alpha}{\gamma} \left( 1 +
\, \frac{\alpha}{\beta} \, \mathcal{S}
\right)^{\beta-1} \neq 1$, it is necessarily $T_\mathrm{G}\neq
T_\mathrm{H}$. Since the Hawking temperature~(\ref{dS6BB}) is the
temperature perceived by observers detecting Hawking radiation, the
generalized temperature $T_\mathrm{G}$ in (\ref{TR1B}) cannot be a
physically meaningful temperature.

In Eq.~(\ref{TR1}), assuming that the thermodynamical energy $E$ is the
black hole mass $M$ leads to an unphysical result. As an
alternative, assume that the thermodynamical temperature $T$ coincides
with the Hawking temperature $T_\mathrm{H}$ instead of assuming $E=M$.
This assumption leads to
\begin{align}
\label{Energy}
dE_\mathrm{G} = T_\mathrm{H} \, d\mathcal{S}_\mathrm{G}
= \frac{d\mathcal{S}_\mathrm{G}}{d\mathcal{S}} \,
\frac{d\mathcal{S}}{\sqrt{16\pi G \mathcal{S}}}
\end{align}
which, in the case of Eq.~(\ref{3-parameter-entropy}), yields
\begin{eqnarray}
\label{Energy2}
dE_\mathrm{G}&=&\frac{\alpha}{\gamma} \left( 1 +
\frac{\alpha}{\beta} \, \mathcal{S} \right)^{\beta-1} \,
\frac{d\mathcal{S}}{\sqrt{16\pi G \mathcal{S}}}\nonumber\\
&=&\frac{\alpha}{\gamma \sqrt{16\pi G}} \left[ \mathcal{S}^{-1/2}
+ \frac{\alpha \left( \beta -1 \right)}{\beta} \, \mathcal{S}^{1/2} +
\mathcal{O}\left( \mathcal{S}^{3/2} \right) \right] \, .
\end{eqnarray}
The integration of Eq.~(\ref{Energy2}) gives
\begin{eqnarray}
\label{GE2}
E_\mathrm{G}&=&\frac{\alpha}{\gamma \sqrt{16\pi G}} \left[ 2
\mathcal{S}^{1/2}
+ \frac{2\alpha \left( \beta -1 \right)}{3\beta} \, \mathcal{S}^{3/2}
+ \mathcal{O}\left( \mathcal{S}^{5/2} \right) \right]\nonumber\\
&=&\frac{\alpha}{\gamma} \left[ M + \frac{4\pi G \alpha \left( \beta -1
\right)}{3\beta} M^3 + \mathcal{O}\left( M^5 \right) \right] \,,
\end{eqnarray}
where the integration constant is determined by the condition that
$E_\mathrm{G}=0$ when $M=0$. Even when $\alpha=\gamma$, due to the
correction $\frac{4\pi G \alpha \left( \beta -1 \right)}{3\beta} M^3$, the
expression~(\ref{GE2}) of the thermodynamical energy $E_\mathrm{R}$
obtained differs from the black hole mass $M$, $E_\mathrm{G}\neq E$, which
seems unphysical.

\section{Cosmology with the 4-parameter generalized entropy} \label{SecI}

Here we consider the 4-parameter generalized entropy (\ref{gen-entropy}), which is indeed more generalized compared to the 3-parameter entropy function 
of Eq.(\ref{3-parameter-entropy}), to describe the cosmological behaviour of the universe 
\cite{Nojiri:2022dkr}. In particular, we examine whether the 4-parameter entropy function 
results to an unified scenario of early inflation and the late dark energy era of the universe. 

The Friedmann-Lema\^{i}tre-Robertson-Walker space-time with flat spacial part will serve our purpose, in particular,
\begin{align}
ds^2=-dt^2+a^2(t)\sum_{i=1,2,3} \left(dx^i\right)^2 \, .
\label{metric}
\end{align}
Here $a(t)$ is called as a scale factor. 

The radius $r_\mathrm{H}$ of the cosmological horizon is given by 
\begin{align}
\label{apphor}
r_\mathrm{H}=\frac{1}{H}\, ,
\end{align}
with $H = \dot{a}/a$ is the Hubble parameter of the universe. 
Then the entropy contained within the cosmological horizon can be obtained from 
the Bekenstein-Hawking relation \cite{Padmanabhan:2009vy}. 
Furthermore the flux of the energy $E$, or equivalently, the increase of the heat $Q$ in the region comes as 
\begin{align}
\label{Tslls2}
dQ = - dE = -\frac{4\pi}{3} r_\mathrm{H}^3 \dot\rho dt = -\frac{4\pi}{3H^3} \dot\rho~dt 
= \frac{4\pi}{H^2} \left( \rho + p \right)~dt \, ,
\end{align}
where, in the last equality, we use the conservation law: $0 = \dot \rho + 3 H \left( \rho + p \right)$. 
Then from the Hawking temperature \cite{Cai:2005ra}
\begin{align}
\label{Tslls6}
T = \frac{1}{2\pi r_\mathrm{H}} = \frac{H}{2\pi}\, ,
\end{align}
and by using the first law of thermodynamics $TdS = dQ$, 
one obtains $\dot H = - 4\pi G \left( \rho + p \right)$. Integrating the expression immediately leads to 
the first FRW equation, 
\begin{align}
\label{Tslls8}
H^2 = \frac{8\pi G}{3} \rho + \frac{\Lambda}{3} \, ,
\end{align}
where the integration constant $\Lambda$ can be treated as a cosmological constant. 

Instead of the Bekenstein-Hawking entropy of Eq.~(\ref{BH-entropy}), we may 
use the generalized entropy in Eq.~(\ref{gen-entropy}), in regard to which, the first law of thermodynamics leads to the following equation:
\begin{align}
\dot{H}\left(\frac{\partial S_\mathrm{g}}{\partial S}\right) = -4\pi G\left(\rho + p\right) \,.
\label{FRW1-sub}
\end{align}
With the explicit form of $S_\mathrm{g}$ from Eq.~(\ref{gen-entropy}), the above equation turns out to be,
\begin{align}
\frac{1}{\gamma}\left[\alpha_{+}\left(1 + \frac{\pi \alpha_+}{\beta GH^2}\right)^{\beta - 1} 
+ \alpha_-\left(1 + \frac{\pi \alpha_-}{\beta GH^2}\right)^{-\beta-1}\right]\dot{H} = -4\pi G\left(\rho + p\right)
\label{FRW-1}
\end{align}
where we use $S = A/(4G) = \pi/(GH^2)$. 
Using the conservation relation of the matter fields, i.e., $\dot{\rho} + 3H\left(\rho + p\right) = 0$, 
Eq.~(\ref{FRW-1}) can be written as,
\begin{align}
\frac{2}{\gamma}\left[\alpha_{+}\left(1 + \frac{\pi \alpha_+}{\beta GH^2}\right)^{\beta - 1} 
+ \alpha_-\left(1 + \frac{\pi \alpha_-}{\beta GH^2}\right)^{-\beta-1}\right]H~dH = \left(\frac{8\pi G}{3}\right)d\rho \,,
 \nonumber
\end{align}
on integrating which, we obtain, 
\begin{align}
\frac{GH^4\beta}{\pi\gamma}&\,\left[ \frac{1}{\left(2+\beta\right)}\left(\frac{GH^2\beta}{\pi\alpha_-}\right)^{\beta}~
2F_{1}\left(1+\beta, 2+\beta, 3+\beta, -\frac{GH^2\beta}{\pi\alpha_-}\right) \right. \nonumber\\ 
&\, \left. + \frac{1}{\left(2-\beta\right)}\left(\frac{GH^2\beta}{\pi\alpha_+}
\right)^{-\beta}~2F_{1}\left(1-\beta, 2-\beta, 3-\beta, -\frac{GH^2\beta}{\pi\alpha_+}\right) \right] = \frac{8\pi G\rho}{3} + \frac{\Lambda}{3} \,,
\label{FRW-2}
\end{align}
where $\Lambda$ is the integration constant (known as the cosmological constant) and $2F_1(\mathrm{arguments})$ denotes the 
Hypergeometric function. Eq.~(\ref{FRW-1}) and Eq.~(\ref{FRW-2}) represent the modified Friedmann 
equations corresponding to the generalized entropy function $S_\mathrm{g}$. 
In the next section, we aim to study 
the cosmological implications of the modified Friedmann Eq.~(\ref{FRW-1}) and Eq.~(\ref{FRW-2}). 

\subsection{Early universe cosmology from the 4-parameter generalized entropy}\label{sec-inf}

During the early stage of the universe we consider the matter field and the cosmological constant ($\Lambda$) to 
be absent, i.e., $\rho = p = \Lambda = 0$. During the early universe, the cosmological constant is highly suppressed with respect to the 
entropic energy density and thus we can safely neglect the $\Lambda$ in studying the early inflationary scenario of the universe. 
Therefore during the early universe, Eq.~(\ref{FRW-2}) becomes,
\begin{align}
& \left[\frac{1}{\left(2+\beta\right)}\left(\frac{GH^2\beta}{\pi\alpha_-}\right)^{\beta}~2F_{1}
\left(1+\beta, 2+\beta, 3+\beta, -\frac{GH^2\beta}{\pi\alpha_-}\right) \right. \nonumber\\ 
&\left. \quad +\frac{1}{\left(2-\beta\right)}\left(\frac{GH^2\beta}{\pi\alpha_+}\right)^{-\beta}~2F_{1}
\left(1-\beta, 2-\beta, 3-\beta, -\frac{GH^2\beta}{\pi\alpha_+}\right) \right] = 0 \, .
 \label{FRW-2-inf}
\end{align}
Here it may be mentioned that the typical 
energy scale during early universe is of the order $\sim 10^{16}\mathrm{GeV}$ ($= 10^{-3}M_\mathrm{Pl}$ where recall that $M_\mathrm{Pl}$ is the Planck mass 
and $M_\mathrm{Pl} = 1/\sqrt{16\pi G}$). This indicates that the condition $GH^2 \ll 1$ holds during the early phase of the universe. 
Owing to such condition, we can safely expand the Hypergeometric function of Eq.~(\ref{FRW-2-inf}) 
as the Taylor series with respect to the argument containing $GH^2$, and as a result, 
the above equation provides a constant Hubble parameter as the solution:
\begin{align}
H = 4\pi M_\mathrm{Pl}\sqrt{\frac{\alpha_+}{\beta}}\left[\frac{(3-\beta)}{(2-\beta)(1-\beta)}\right] \,.
\label{dS Hubble parameter-sol}
\end{align}
For $\frac{\alpha_+}{\beta} \sim 10^{-6}$ and $\beta \lesssim \mathcal{O}(1)$, the constant Hubble parameter can be fixed at 
$H \sim 10^{-3}M_\mathrm{Pl}$ which can be identified with typical inflationary energy scale. 
Therefore the entropic cosmology corresponding to the 
generalized entropy function $S_\mathrm{g}$ leads to a de-Sitter inflationary scenario during the early universe. 
However, a de-Sitter inflation has no exit mechanism, and moreover, 
the primordial curvature perturbation gets exactly 
scale invariant in the context of a de-Sitter inflation, which is not consistent with the recent Planck data \cite{Akrami:2018odb} at all. 
This indicates that the constant 
Hubble parameter obtained in Eq.~(\ref{dS Hubble parameter-sol}) does not lead to a good inflationary scenario of the universe. 
Thus in order to achieve a viable quasi de-Sitter inflation in the present context, 
we consider the parameters of $S_\mathrm{g}$ to be slowly varying functions 
with respect to the cosmic time. In particular, we consider the parameter 
$\gamma$ to vary and the other parameters (i.e., $\alpha_+$, $\alpha_-$ and $\beta$) remain constant with $t$. In particular,
\begin{align}
\gamma(N)=\left\{ 
\begin{array}{ll}
\gamma_0~\exp{\left[-\int_{N}^{N_f}\sigma(N)~dN\right]}\quad &;\ N \leq N_f \\
\gamma_0 &;\ N \geq N_f \,,
\end{array}
\right. 
\label{gamma function}
\end{align}
where $\gamma_0$ is a constant and $N$ denotes the inflationary e-folding number with $N_f$ being the total e-folding number of the inflationary era. 
The function $\sigma(N)$ has the following form,
\begin{align}
\sigma(N) = \sigma_0 + \mathrm{e}^{-\left(N_f - N\right)} \,,
\label{sigma function}
\end{align}
where $\sigma_0$ is a constant. 
The second term in the expression of $\sigma(N)$ becomes effective only when $N\approx N_f$, i.e.,  near the end of inflation. 
The term $\mathrm{e}^{-\left(N_f - N\right)}$ in Eq.~(\ref{sigma function}) is actually considered to ensure an exit 
from inflation era and thus proves 
to be an useful one to make the inflationary scenario viable. In such scenario where $\gamma$ varies with $N$, the Friedmann equation 
turns out to be, 
\begin{align}
 -\left(\frac{2\pi}{G}\right)
\left[\frac{\alpha_+\left(1 + \frac{\alpha_+}{\beta}~S\right)^{\beta-1} + \alpha_-\left(1 + \frac{\alpha_-}{\beta}~S\right)^{-\beta-1}}
{\left(1 + \frac{\alpha_+}{\beta}~S\right)^{\beta} - \left(1 + \frac{\alpha_-}{\beta}~S\right)^{-\beta}}\right]\frac{H'(N)}{H^3} = \sigma(N) \,.
\label{FRW-eq-viable-inf-main1}
\end{align}
By using $S = \pi/(GH^2)$, or equivalently, $2HdH = -\frac{\pi}{GS^2}dS$, one can integrate Eq.(\ref{FRW-eq-viable-inf-main1}) to get $H(N)$ as,
\begin{align}
H(N) = 4\pi M_\mathrm{Pl}\sqrt{\frac{\alpha_+}{\beta}}
\left[\frac{2^{1/(2\beta)}\exp{\left[-\frac{1}{2\beta}\int_0^{N}\sigma(N)dN\right]}}
{\left\{1 + \sqrt{1 + 4\left(\alpha_+/\alpha_-\right)^{\beta}\exp{\left[-2\int_0^{N}\sigma(N)dN\right]}}\right\}^{1/(2\beta)}}\right] \,.
\label{solution-viable-inf-2}
\end{align}
The above solution of $H(N)$ allows an exit from inflation at finite e-fold number which can be fixed at $N_\mathrm{f}= 58$ for suitable choices of 
the entropic parameters \cite{Nojiri:2022dkr}. Moreover we determine the spectral index for curvature perturbation ($n_s$) and the tensor-to-scalar ratio ($r$) in the present context of entropic 
cosmology, and they are given by \cite{Nojiri:2022dkr}: 
\begin{align}
n_s = 1 - \frac{2\sigma_0\sqrt{1 + 4\left(\alpha_+/\alpha_-\right)^{\beta}\exp{\left[-2\left(1 + \sigma_0N_f\right)\right]}}}
{(1+\sigma_0)\sqrt{1 + 4\left(\alpha_+/\alpha_-\right)^{\beta}}} - \frac{8\sigma_0\left(\alpha_+/\alpha_-\right)^{\beta}}
{1 + 4\left(\alpha_+/\alpha_-\right)^{\beta}} \,,
\label{ns final form}
\end{align}
and 
\begin{align}
r = \frac{16\sigma_0\sqrt{1 + 4\left(\alpha_+/\alpha_-\right)^{\beta}\exp{\left[-2\left(1 + \sigma_0N_f\right)\right]}}}
{(1+\sigma_0)\sqrt{1 + 4\left(\alpha_+/\alpha_-\right)^{\beta}}}
\label{r final form}
\end{align}
respectively. It turns out that the theoretical expectations of $n_s$ and $r$ get simultaneously compatible with the Planck data for the following ranges 
of the parameters:
\begin{align}
\sigma_0=&\,[0.013,0.017] \,, \quad \left(\alpha_+/\alpha_-\right)^{\beta} \geq 7.5 \,,\nonumber\\
\beta=&\, (0,0.4]\ \mbox{and}\ \left(\alpha_+/\beta\right) \approx 10^{-6} \,,
\label{inf-constraints}
\end{align}
for $N_f = 58$. The consideration of $\frac{\alpha_+}{\beta} \sim 10^{-6}$ leads to the energy scale at the onset 
of inflation as $H \sim 10^{-3}M_\mathrm{Pl}$.

\subsection{Dark energy era from the 4-parameter generalized entropy}\label{sec-DE}

\begin{figure}[!h]
\begin{center}
\centering
\includegraphics[width=3.5in,height=2.5in]{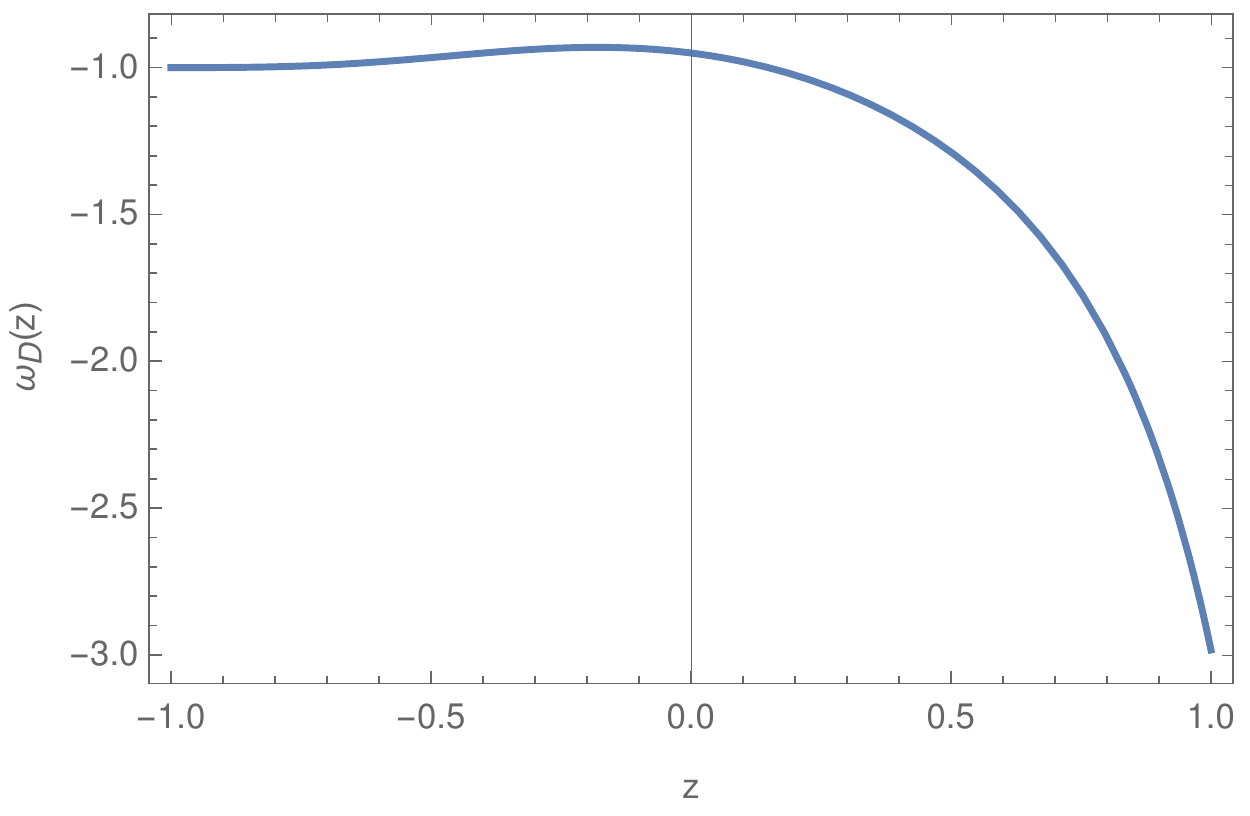}
\caption{$\omega_\mathrm{D}(z)$ vs. $z$ for a particular set of values of the parameters from their viable ranges 
as per Eq.(\ref{inf-constraints}) and Eq.(\ref{constraint-gamma-1}), say 
$\beta = 0.35$, $\left(\alpha_+/\alpha_-\right)^{\beta} = 10$, $\alpha_{+}/\beta = 10^{-6}$ and $\gamma_m = 1.8\times10^{-4}$.} 
\label{plot-eos1}
\end{center}
\end{figure}

In this section we will concentrate on late time cosmological implications of the generalized entropy function ($S_\mathrm{g}$), 
where the cosmological constant $\Lambda$ is considered to be non-zero. 
During the late time, the parameter $\gamma$ becomes constant, in particular $\gamma = \gamma_0$, as we demonstrated in Eq.~(\ref{gamma function}). 
As a result, the entropy function at the late time takes the following form,
\begin{align}
S_\mathrm{g} = \frac{1}{\gamma_0}\left[\left(1 + \frac{\alpha_+}{\beta}~S\right)^{\beta} - \left(1 + \frac{\alpha_-}{\beta}~
S\right)^{-\beta}\right] \,,
\label{entropy-late time}
\end{align}
with $S = \pi/(GH^2)$. 
Consequently, the energy density and pressure corresponding to the $S_\mathrm{g}$ are given by,
\begin{align}
\rho_\mathrm{g}=&\, \frac{3H^2}{8\pi G}\left[1 - \frac{\alpha_+}{\gamma_0(2-\beta)}\left(\frac{GH^2\beta}{\pi\alpha_+}\right)^{1-\beta}\right] \,,
\nonumber\\
p_\mathrm{g}=&\, -\frac{\dot{H}}{4\pi G}\left[1 - \frac{\alpha_+}{\gamma_0}\left(\frac{GH^2\beta}{\pi\alpha_+}\right)^{1-\beta} 
 - \left(\frac{\alpha_+}{\gamma_0}\right)\left(\frac{\alpha_+}{\alpha_-}\right)^{\beta}\left(\frac{GH^2\beta}{\pi\alpha_+}\right)^{1+\beta}\right] 
 - \rho_\mathrm{g} \,.
\label{energy and pressure-late time}
\end{align}
Therefore the dark energy density ($\rho_\mathrm{D}$) 
is contributed from the entropic energy density ($\rho_\mathrm{g}$) as well as from the cosmological constant. 
In particular
\begin{eqnarray}
 \rho_\mathrm{D}&=&\rho_\mathrm{g} + \frac{3}{8\pi G}\left(\frac{\Lambda}{3}\right)~~,\nonumber\\
 \rho_\mathrm{D} + p_\mathrm{D}&=&\rho_\mathrm{g} + p_\mathrm{g}~~.
 \label{de-1}
\end{eqnarray}
Consequently, the dark energy EoS parameter comes with the following expression:
\begin{align}
\omega_\mathrm{D} = p_\mathrm{D}/\rho_\mathrm{D} = -1 - \left(\frac{2\dot{H}}{3H^2}\right)
\left[\frac{1 - \frac{\alpha_+}{\gamma_0}\left(\frac{GH^2\beta}{\pi\alpha_+}\right)^{1-\beta} 
 - \left(\frac{\alpha_+}{\gamma_0}\right)\left(\frac{\alpha_+}{\alpha_-}\right)^{\beta}\left(\frac{GH^2\beta}{\pi\alpha_+}\right)^{1+\beta}}
{1 - \frac{\alpha_+}{\gamma_0(2-\beta)}\left(\frac{GH^2\beta}{\pi\alpha_+}\right)^{1-\beta} + \frac{\Lambda}{3H^2}}\right] \,.
\label{eos-1}
\end{align}
In presence of the cosmological constant, the Friedmann equations are written as,
\begin{align}
H^2 = \frac{8\pi G}{3}\left(\rho_m + \rho_\mathrm{D}\right) 
= \frac{8\pi G}{3}\left(\rho_m + \rho_\mathrm{g}\right) + \frac{\Lambda}{3} \,,\nonumber\\
\dot{H} = -4\pi G\left[\rho_m + \left(\rho_\mathrm{D} + p_\mathrm{D}\right)\right] 
= -4\pi G\left[\rho_m + \left(\rho_\mathrm{g} + p_\mathrm{g}\right)\right] \,.
\label{FRW-late time-1}
\end{align} 
As usual, the fractional energy density of the pressureless matter and the dark energy satisfy $\Omega_m + \Omega_\mathrm{D} = 1$ which along with 
$\rho_m = \rho_{m0}\left(\frac{a_0}{a}\right)^3$ (with $\rho_{m0}$ being 
the present matter energy density) result to the Hubble parameter in terms of the red shift factor ($z$) as follows,
\begin{align}
H(z) = \frac{H_0\sqrt{\Omega_{m0}(1+z)^3}}{\sqrt{1-\Omega_\mathrm{D}}} \,.
\label{Hubble late time-1}
\end{align}
Plugging the expression of $\rho_\mathrm{g}$ from Eq.~(\ref{energy and pressure-late time}) into 
$\Omega_\mathrm{D} = \left(\frac{8\pi G}{3H^2}\right)\rho_\mathrm{g} + \frac{\Lambda}{3}$, and using the above form of $H(z)$, we obtain,
\begin{align}
\Omega_\mathrm{D}(z) = 1 - \frac{\left[\frac{\alpha_+}{\gamma_0(2-\beta)}\right]^{\frac{1}{2-\beta}}
\left[\frac{GH_0^2\beta}{\pi\alpha_+}~\Omega_{m0}(1+z)^3\right]^{\frac{1-\beta}{2-\beta}}}
{\left[1 + \frac{\Lambda}{3H_0^2\Omega_{m0}(1+z)^3}\right]^{1/(2-\beta)}}\,.
\label{fractional DE late time-1}
\end{align}  
By using the above expressions, we determine the DE EoS parameter from Eq.~(\ref{eos-1}) as follows (see \cite{Nojiri:2022dkr}), 
\begin{align}
\omega_\mathrm{D}(z) = -1 + \frac{1}{(2-\beta)\left(1 + \frac{\Lambda}{3H_0^2\Omega_{m0}(1+z)^3}\right)}\left(\frac{N}{D}\right) \,,
\label{eos final-1}
\end{align}
where $N$ (the numerator) and $D$ (the denominator) have the following forms,
\begin{eqnarray}
N=1 - \Omega_{m0}(2-\beta)\left(1+z\right)^{\frac{3(1-\beta)}{(2-\beta)}}
\Bigg\{\left(\frac{1+\frac{\Lambda}{3H_0^2\Omega_{m0}}}{\left[f(\Lambda,\Omega_{m0},H_0,z)\right]^{1-\beta}}\right) 
&+&\left(\frac{\alpha_+}{\alpha_-}\right)^{\beta}\left[\Omega_{m0}(2-\beta)\gamma_0/\alpha_+\right]^{\frac{2\beta}{1-\beta}}
\left(1 + z\right)^{\frac{6\beta}{2-\beta}}\nonumber\\
&\times&\left(\frac{\left[1+\frac{\Lambda}{3H_0^2\Omega_{m0}}\right]^{(1+\beta)/(1-\beta)}}
{\left[f(\Lambda,\Omega_{m0},H_0,z)\right]^{1+\beta}}\right)\Bigg\} \,,
\nonumber
\end{eqnarray}
and
\begin{eqnarray}
D = 1 - \Omega_{m0}\left(1+z\right)^{\frac{3(1-\beta)}{(2-\beta)}}\left(\frac{1+\frac{\Lambda}{3H_0^2\Omega_{m0}}}
{\left[f(\Lambda,\Omega_{m0},H_0,z)\right]^{1-\beta}}\right) + \frac{\Lambda}{3H_0^2}\left(\frac{f(\Lambda,\Omega_{m0},H_0,z)}
{(1+z)^{3/(2-\beta)}}\right)
\label{N and D-1}
\end{eqnarray}
respectively. Therefore $\omega_\mathrm{D}$ depends on the parameters: 
$\beta$, $\left(\alpha_+/\alpha_-\right)^{\beta}$, $\gamma_0$ and $\alpha_+$. Recall that 
the inflationary quantities are found to be simultaneously compatible with the Planck data if some of the parameters 
like $\alpha_{+}$, $\alpha_{-}$ and $\beta$ get constrained according to Eq.~(\ref{inf-constraints}), while the parameter $\gamma_0$ remains 
free from the inflationary requirement. With the aforementioned ranges of $\alpha_+$, $\alpha_-$ and $\beta$, $\omega_\mathrm{D}(0)$ 
becomes compatible with the Planck observational data, 
provided $\gamma_0$ lies within a small window as follows,
\begin{align}
1.5\times10^{-4} \leq \frac{\gamma_0}{\left(8\pi GH_0^2\right)^{1-\beta}} \leq 2\times10^{-4} \,.\label{constraint-gamma-1}
\end{align}
Furthermore the deceleration parameter (symbolized by $q$) at present universe is obtained as,
\begin{eqnarray}
 q = -1 + \frac{3}{2(2-\beta)\left(1 + \frac{\Lambda}{3H_0^2\Omega_{m0}}\right)}~~.
 \label{q}
\end{eqnarray}
Therefore for $\gamma_m = [1.5\times10^{-4},2\times10^{-4}]$, the theoretical expression of $q$ lies within 
$q = [-0.56,-0.42]$ which certainly contains the observational value of $q = -0.535$ from the Planck data \cite{Planck:2018vyg}. In particular, 
$q = -0.535$ occurs for $\gamma_m = 1.8\times10^{-4}$. Considering this value of $\gamma_m$ and by using 
Eq.(\ref{eos final-1}), we give the plot of $\omega_\mathrm{D}(z)$ vs. $z$, see Fig.~\ref{plot-eos1}. 
The figure reveals that that the theoretical expectation of the 
DE EoS parameter at present time acquires the value: $\omega_\mathrm{D}(0) = -0.950$ which is well consistent with the Planck observational 
data \cite{Planck:2018vyg}.
 
As a whole, we may argue that the entropic cosmology from the generalized entropy function $S_\mathrm{g}$ can 
unify the early inflation to the late 
dark energy era of the universe, for suitable ranges of the parameters given by:
\begin{align}
\sigma_0=&\,[0.013,0.017] \,, \quad \left(\alpha_+/\alpha_-\right)^{\beta} \geq 7.5 \,,\nonumber\\
\beta=&\, (0,0.4]\ \mbox{and}\ \gamma_m = [1.5\times10^{-4},2\times10^{-4}] \,.
\label{constraints-final}
\end{align}

Despite these successes, here it deserves mentioning that the entropy function $S_\mathrm{g}$ 
seems to be plagued with singularity for certain cosmological evolution of the universe, 
in particular, in the context of bounce cosmology. Due to the reason that the Bekenstein-Hawking entropy can be expressed as 
$S = \pi/\left(GH^2\right)$, the generalized entropy $S_\mathrm{g}$ contains factor that is proportional to $1/H^2$ which diverges 
at $H = 0$, for instance at the instant of bounce in the context of bounce cosmology. 
Therefore in a bounce scenario, the generalized entropy function shown 
in Eq.(\ref{gen-entropy}) is not physical, and thus, we need to search for a different generalized entropy function which can lead 
to various known entropy functions for suitable choices of the parameters, and at the same time, 
proves to be non-singular for the entire cosmological evolution of the universe even at $H = 0$.

\section{Search for a singular-free generalized entropy}\label{sec-gen-entropy}

With this spirit, we propose a new singular-free entropy function given by \cite{OP-submitted},
\begin{eqnarray}
S_\mathrm{ns}\left[\alpha_{\pm},\beta,\gamma,\epsilon\right] = \frac{1}{\gamma}\bigg[\left\{1 + \frac{1}{\epsilon}\tanh\left(\frac{\epsilon \alpha_+}{\beta}~S\right)\right\}^{\beta} 
- \left\{1 + \frac{1}{\epsilon}\tanh\left(\frac{\epsilon \alpha_-}{\beta}~S\right)\right\}^{-\beta}\bigg]~~,
\label{gen-entropy-b}
\end{eqnarray}
where $\alpha_{\pm}$, $\beta$, $\gamma$ and $\epsilon$ are the parameters which are considered to be positive, $S$ symbolizes 
the Bekenstein-Hawking entropy and the suffix 'ns' stands for 'non-singular'. 
In regard to the number of parameters, we propose a conjecture at the end of this section. 
First we demonstrate that the above entropy function remains finite, and thus is non-singular, during the 
whole cosmological evolution of a bouncing universe. In particular, 
the $S_\mathrm{g}$ takes the following form at the instant of bounce:
\begin{eqnarray}
S_\mathrm{ns}\left[\alpha_{\pm},\beta,\gamma,\epsilon\right] = 
\frac{1}{\gamma}\bigg[\left\{1 + \frac{1}{\epsilon}\right\}^{\beta} 
- \left\{1 + \frac{1}{\epsilon}\right\}^{-\beta}\bigg]~~.
\label{gen-entropy-2}
\end{eqnarray}
Having demonstrated the non-singular behaviour of the entropy function, we now show that $S_\mathrm{ns}$ of Eq.(\ref{gen-entropy-b}), 
for suitable choices of the parameters, 
reduces to various known entropies proposed so far. 

\begin{itemize}
\item For $\epsilon \rightarrow 0$, $\alpha_+ \rightarrow \infty$ and $\alpha_- = 0$ along with the identification 
$\gamma = \left(\alpha_+/\beta\right)^{\beta}$, $S_\mathrm{ns}$ converges to the Tsallis entropy or to the Barrow entropy respectively.
\item The limit $\epsilon \rightarrow 0$, $\alpha_- = 0$, $\beta \rightarrow 0$ and $\frac{\alpha_+}{\beta} \rightarrow \mathrm{finite}$ 
results to the similarity between the non-singular generalized entropy $S_\mathrm{g}$ and the R\'{e}nyi entropy.
\item For $\epsilon \rightarrow 0$ and $\alpha_- \rightarrow 0$, the non-singular generalized entropy converges to the following form,
\begin{eqnarray}
 S_\mathrm{ns} = \frac{1}{\gamma}\left[\left(1 + \frac{\alpha_+}{\beta}~S\right)^{\beta} - 1\right]\label{SM}
\end{eqnarray}
Therefore with $\gamma = R$, $\alpha_+ = R$ and $\beta = R/\delta$, the above form of $S_\mathrm{ns}$ becomes similar to the 
Sharma-Mittal entropy.

\item For $\epsilon \rightarrow 0$, 
$\beta \rightarrow \infty$, $\alpha_+ = \alpha_- = \frac{\gamma}{2} = K$ -- the generalized entropy converges to the form of 
Kaniadakis entropy,
\item Finally, $\epsilon \rightarrow 0$, $\alpha_- \rightarrow 0$, $\beta \rightarrow \infty$ 
and $\gamma = \alpha_+ = (1-q)$, the generalized entropy of Eq.~(\ref{gen-entropy-b}) gets resemble with the Loop Quantum Gravity entropy.

\end{itemize}

Furthermore, the generalized entropy function in Eq.~(\ref{gen-entropy-b}) shares the following properties: 
(1) the non-singular generalized entropy 
satisfies the generalized third law of thermodynamics. (2) 
$S_\mathrm{ns}\left[ \alpha_{\pm},\beta,\gamma,\epsilon \right]$ turns out to be a monotonically increasing function of $S$. 
(3) $S_\mathrm{ns}\left[ \alpha_{\pm},\beta,\gamma,\epsilon \right]$ proves to converge to the Bekenstein-Hawking 
entropy at certain limit of the parameters.

At this stage it deserves mentioning that we have proposed two different generalized entropy functions in Eq.(\ref{gen-entropy}) and 
in Eq.(\ref{gen-entropy-b}) respectively -- the former entropy function contains four independent parameters while the latter one 
has five parameters. Furthermore both the entropies are able to generalize the known entropies 
for suitable choices of the respective parameters. However as mentioned earlier that the entropy with four parameters becomes singular 
at $H = 0$ (for instance, in a bounce scenario when the Hubble parameter vanishes at the instant of bounce), while the entropy function having five 
parameters proves to be singular-free during the whole cosmological evolution of the universe. Based on these findings, we give a second conjecture 
regarding the number of parameters in the non-singular generalized entropy function:\\

\textbf{\underline{Conjecture - II}}: 
``The minimum number of parameters required in a generalized entropy function that can generalize 
all the known entropies, and at the same time, is also singular-free during the universe's evolution -- 
is equal to five''.

\section{Cosmology with the non-singular generalized entropy} \label{SecI-bounce}

Applying the thermodynamic laws to the non-singular generalized entropy function $S_\mathrm{ns}$ and by following the same 
procedure as of Sec.[\ref{SecI}], one gets the cosmological field equations corresponding to the $S_g\mathrm{ns}$ \cite{OP-submitted}: 
\begin{eqnarray}
\frac{1}{\gamma}&\bigg[&\alpha_{+}~\mathrm{sech}^2\left(\frac{\epsilon\pi \alpha_+}{\beta GH^2}\right)
\left\{1 + \frac{1}{\epsilon}\tanh\left(\frac{\epsilon\pi \alpha_+}{\beta GH^2}\right)\right\}^{\beta-1}\nonumber\\
&+&\alpha_{-}~\mathrm{sech}^2\left(\frac{\epsilon\pi \alpha_-}{\beta GH^2}\right)
\left\{1 + \frac{1}{\epsilon}\tanh\left(\frac{\epsilon\pi \alpha_-}{\beta GH^2}\right)\right\}^{-\beta-1}\bigg]\dot{H} = -4\pi G\left(\rho + p\right)~~.
\label{FRW-1-b}
\end{eqnarray}
Owing to the conservation equation of matter fields, in particular $\dot{\rho} + 3H\left(\rho + p\right) = 0$, 
the above expression can be integrated to get
\begin{align}
f\left(H;~\alpha_{\pm},\beta,\gamma,\epsilon\right) = \frac{8\pi G\rho}{3} + \frac{\Lambda}{3} \,.
\label{FRW-2-b}
\end{align}
Here the integration constant is symbolized by $\Lambda$ and the function $f$ has the following form:
\begin{eqnarray}
 f\left(H;~\alpha_{\pm},\beta,\gamma,\epsilon\right) = 
 \frac{2}{\gamma}\int&\bigg[&\alpha_{+}~\mathrm{sech}^2\left(\frac{\epsilon\pi \alpha_+}{\beta GH^2}\right)
\left\{1 + \frac{1}{\epsilon}\tanh\left(\frac{\epsilon\pi \alpha_+}{\beta GH^2}\right)\right\}^{\beta-1}\nonumber\\
&+&\alpha_{-}~\mathrm{sech}^2\left(\frac{\epsilon\pi \alpha_-}{\beta GH^2}\right)
\left\{1 + \frac{1}{\epsilon}\tanh\left(\frac{\epsilon\pi \alpha_-}{\beta GH^2}\right)\right\}^{-\beta-1}\bigg]H~dH~~.
\label{f}
\end{eqnarray}
In regard to the functional form of $f\left(H;~\alpha_{\pm},\beta,\gamma,\epsilon\right)$, we would like to mention that the integration 
in Eq.(\ref{f}) may not be performed in a closed form, unless certain conditions are imposed. 
For example, we consider $GH^2 \ll 1$ which is, in fact, valid during the 
universe's evolution (i.e the Hubble parameter is less than the Planck scale). With $GH^2 \ll 1$, the functional form of $f$ turns out to be,
\begin{eqnarray}
 f\left(H;~\alpha_{\pm},\beta,\gamma,\epsilon\right) = 
 \frac{4}{\gamma}H^2&\bigg\{&\alpha_{+}\left(1+\frac{1}{\epsilon}\right)^{\beta-1}\left[\mathrm{exp}\left(-\frac{2\epsilon\pi \alpha_+}{\beta GH^2}\right) 
 + \left(\frac{2\epsilon\pi \alpha_+}{\beta GH^2}\right)\mathrm{Ei}\left(-\frac{2\epsilon\pi \alpha_+}{\beta GH^2}\right)\right]\nonumber\\
 &+&\alpha_{-}\left(1+\frac{1}{\epsilon}\right)^{-\beta-1}\left[\mathrm{exp}\left(-\frac{2\epsilon\pi \alpha_-}{\beta GH^2}\right) 
 + \left(\frac{2\epsilon\pi \alpha_-}{\beta GH^2}\right)\mathrm{Ei}\left(-\frac{2\epsilon\pi \alpha_-}{\beta GH^2}\right)\right]\bigg\}~~.
 \label{f-aprroximated-2}
\end{eqnarray}
Therefore as a whole, Eq.~(\ref{FRW-1-b}) and Eq.~(\ref{FRW-2-b}) are the cosmological field equations 
corresponding to the generalized entropy $S_\mathrm{g}$. 

\subsection{Non-singular entropy on bounce cosmology}\label{sec-bounce}

In this section, we will address the implications of the generalized entropy $S_\mathrm{ns}$ on non-singular bounce cosmology, in particular, 
we will investigate whether the entropic energy density can trigger a viable bounce during the early stage of the universe that is consistent 
with the observational constraints. For this purpose, we take 
the matter field and the cosmological constant to be absent, i.e., $\rho = p = \Lambda = 0$. In effect, Eq.~(\ref{FRW-1-b}) becomes,
\begin{eqnarray}
\frac{1}{\gamma}&\bigg[&\alpha_{+}~\mathrm{sech}^2\left(\frac{\epsilon\pi \alpha_+}{\beta GH^2}\right)
\left\{1 + \frac{1}{\epsilon}\tanh\left(\frac{\epsilon\pi \alpha_+}{\beta GH^2}\right)\right\}^{\beta-1}\nonumber\\
&+&\alpha_{-}~\mathrm{sech}^2\left(\frac{\epsilon\pi \alpha_-}{\beta GH^2}\right)
\left\{1 + \frac{1}{\epsilon}\tanh\left(\frac{\epsilon\pi \alpha_-}{\beta GH^2}\right)\right\}^{-\beta-1}\bigg]\dot{H} = 0~~.
\label{FRW-1-bounce}
\end{eqnarray}
The parameters $\left(\alpha_{\pm},\beta,\gamma,\epsilon\right)$ are positive, and thus the solution of the above equation is given by: 
$\dot{H} = 0$ or equivalently $H=\mathrm{constant}$. Clearly 
$H=\mathrm{constant}$ does not lead to the correct evolution of the universe. Thus similar to the previous case, 
we consider the parameters of $S_\mathrm{ns}[\alpha_{\pm},\beta,\gamma,\epsilon]$ vary with time. 
In particular, we consider the parameter $\gamma$ to vary with time, and all the other parameters remain fixed, i.e.
\begin{eqnarray}
 \gamma = \gamma(N)~~,
 \label{gamma}
\end{eqnarray}
with $N$ being the e-fold number of the universe. In such scenario where $\gamma(N)$ varies with time, the Friedmann equation corresponds to 
$S_\mathrm{ns}[\alpha_{\pm},\beta,\gamma,\epsilon]$ gets modified compared to Eq.(\ref{FRW-1-bounce}), and is given by:
\begin{eqnarray}
\left[\frac{\alpha_{+}~\mathrm{sech}^2\left(\frac{\epsilon \alpha_+}{\beta}S\right)
\left\{1 + \frac{1}{\epsilon}\tanh\left(\frac{\epsilon\alpha_+}{\beta}S\right)\right\}^{\beta-1} 
+ \alpha_{-}~\mathrm{sech}^2\left(\frac{\epsilon\alpha_-}{\beta}S\right)
\left\{1 + \frac{1}{\epsilon}\tanh\left(\frac{\epsilon\alpha_-}{\beta}S\right)\right\}^{-\beta-1}}
{\left\{1 + \frac{1}{\epsilon}\tanh\left(\frac{\epsilon\alpha_+}{\beta}S\right)\right\}^{\beta} - 
\left\{1 + \frac{1}{\epsilon}\tanh\left(\frac{\epsilon\alpha_-}{\beta}S\right)\right\}^{-\beta}}\right]dS = \frac{\gamma'(N)}{\gamma(N)}dN
\label{bounce-3}
\end{eqnarray}
where an overprime denotes $\frac{d}{d\eta}$. Eq.(\ref{bounce-3}) can be integrated to get,
\begin{eqnarray}
 \tanh{\left(\frac{\epsilon \pi\alpha}{\beta GH^2}\right)} = \left\{\frac{\gamma(N) + \sqrt{\gamma^2(N) + 4}}{2}\right\}^{1/\beta} - 1~~.
 \label{bounce-6}
\end{eqnarray}
where we take $\alpha_+ = \alpha_- = \alpha$ (say, without losing any generality) in order to extract an explicit solution of $H(N)$. 
Due to the appearance of quadratic power of $H$, Eq.(\ref{bounce-6}) allows a positive branch as well as a negative branch of the Hubble parameter. 
This leads to a natural possibility of symmetric bounce in the present context of singular free generalized entropic cosmology. Moreover Eq.(\ref{bounce-6}) 
also demonstrates that the explicit evolution of $H(N)$ does depend on the form of $\gamma(N)$. In the following, we will consider two cases where 
we will determine the form of $\gamma(N)$ such that it gives two different symmetric bounce scenarios respectively. 
\begin{enumerate}
 \item The exponential bounce described by the scale factor,
 \begin{eqnarray}
 a(t) = \mathrm{exp}\left(a_0t^2\right)~~.
 \label{exp bounce-1}
\end{eqnarray}
This results to a symmetric bounce at $t = 0$. Here $a_0$ is a constant having mass dimension [+2] -- this constant is related 
with the entropic parameters of $S_\mathrm{ns}$ and thus, without losing any generality, 
we take $a_0 = \frac{\epsilon \pi\alpha}{4G\beta}$. Such an exponential bounce can be achieved from singular free entropic cosmology provided the 
$\gamma(N)$ is given by,
\begin{eqnarray}
 \gamma(N) = \left\{1 + \frac{1}{\epsilon}\tanh\left(\frac{1}{N}\right)\right\}^{\beta} - 
\left\{1 + \frac{1}{\epsilon}\tanh\left(\frac{1}{N}\right)\right\}^{-\beta}~~.
\label{exp bounce-4}
\end{eqnarray}
\item The quasi-matter bounce is described by,
In this case, the scale factor is,
\begin{eqnarray}
 a(t) = \left[1 + a_0\left(\frac{t}{t_0}\right)^2\right]^n
 \label{matter bounce-1}
\end{eqnarray}
which is symmetric about $t = 0$ when the bounce happens. The $n$, $a_0$ and $t_0$ considered in the scale factor are related to 
the entropic parameters, and we take it as follows:
\begin{eqnarray}
 n = \sqrt{\alpha}~~~~~~,~~~~~~~~a_0 = \frac{\pi}{4\beta}~~~~~~~~\mathrm{and}~~~~~~~~~t_0 = \sqrt{G/\epsilon}~~,
 \label{matter bounce-2}
\end{eqnarray}
with $G$ being the gravitational constant. The relation between ($n$, $a_0$, $t_0$) with the entropic parameters can be considered in a different 
way compared to the Eq.(\ref{matter bounce-2}), however for a simplified expression of $\gamma(N)$ we consider the relations as of 
Eq.(\ref{matter bounce-2}). Consequently the $\gamma(N)$ which leads to such quasi-matter bounce, comes as,
\begin{eqnarray}
\gamma(N) = \left\{1 + \frac{1}{\epsilon}\tanh\left[\mathrm{e}^{-N/\sqrt{\alpha}}\left(\mathrm{e}^{N/\sqrt{\alpha}} - 1\right)^{\frac{1}{2}}\right]\right\}^{\beta} - 
\left\{1 + \frac{1}{\epsilon}\tanh\left[\mathrm{e}^{-N/\sqrt{\alpha}}\left(\mathrm{e}^{N/\sqrt{\alpha}} - 1\right)^{\frac{1}{2}}\right]\right\}^{-\beta}~~.
\label{matter bounce-6}
\end{eqnarray}
\end{enumerate}
Here it deserves mentioning that in the case of exponential bounce, the comoving Hubble radius asymptotically goes to zero and thus the perturbation 
modes remain at the super-Hubble regime at the distant past. This may results to the ``horizon problem'' in the exponential bounce scenario. 
On contrary, the comoving Hubble radius in the case of quasi-matter bounce asymptotically diverges to infinity at both sides of the bounce, 
and thus the perturbation modes lie within the deep sub-Hubble regime at the distant past -- this resolves the horizon issue. Based on this arguments, 
we will concentrate on the quasi-matter bounce to perform the perturbation analysis.

In regard to the perturbation analysis, we represent the present entropic cosmology with the ghost free Gauss-Bonnet (GB) theory of gravity 
proposed in \cite{Nojiri:2018ouv}. The motivation of such representation is due to the rich structure of the Gauss-Bonnet theory in various 
directions of cosmology \cite{Odintsov:2022unp,Elizalde:2020zcb,Bamba:2020qdj,Nojiri:2022xdo}. The 
action for $f(\mathcal{G})$ gravity is given by \cite{Nojiri:2018ouv},
\begin{equation}
\label{FRGBg19} S=\int d^4x\sqrt{-g} \left(\frac{1}{2\kappa^2}R 
+ \lambda \left( \frac{1}{2} \partial_\mu \chi \partial^\mu \chi 
+ \frac{\mu^4}{2} \right) - \frac{1}{2} \partial_\mu \chi \partial^\mu \chi
+ h\left( \chi \right) \mathcal{G} - V\left( \chi \right)\right)\, ,
\end{equation}
where $\mu$ is a constant having mass dimension $[+1]$, $\lambda$ represents the Lagrange multiplier, $\chi$ is a scalar field and $V(\chi)$ is its 
potential. Moreover $\mathcal{G} = R^2 - 4R_{\mu\nu}R^{\mu\nu} + R_{\mu\nu\alpha\beta}R^{\mu\nu\alpha\beta}$ is the Gauss-Bonnet scalar and 
$h(\chi)$ symbolizes the Gauss-Bonnet coupling with the scalar field. Moreover we consider such class of Gauss-Bonnet coupling functions 
that satisfy $\ddot{h} = \dot{h}H$. This condition actually leads to the speed of the gravitational wave as unity in the context of 
GB theory and makes the model compatible with the GW170817 event. For a 
certain $\gamma(N)$ in the context of entropic cosmology, there exists an equivalent set of GB parameters in the side of Gauss-Bonnet cosmology that 
results to the same cosmological evolution as of the generalized entropy. In particular, the equivalent forms of 
$\tilde{V}(\chi)$ and $\lambda(t)$ for a certain $\gamma(N)$ turn out to be,
\begin{eqnarray}
\tilde{V}(\chi)&=&-8\pi G~F_1\left[\gamma(N),\gamma'(N)\right]\left(\frac{1}{\kappa^2} + 8h_0a(t)H(t)\right) \bigg|_{t=\chi/\mu^2}\, ,
\label{equiv potential}\\
\mu^4\lambda(t)&=&-8\pi G~F_2\left[\gamma(N),\gamma'(N)\right]\left(\frac{1}{\kappa^2} - 8h_0a(t)H(t)\right)\, ,
\label{equiv potential and LM}
\end{eqnarray}
where the functions $F_1\left[\gamma(N),\gamma'(N)\right]$ and $F_2\left[\gamma(N),\gamma'(N)\right]$ are given by,
\begin{eqnarray}
 F_1\left[\gamma(N),\gamma'(N)\right]&=&-\left(\frac{3\epsilon\alpha}{4\beta G^2}\right)
 \left[\ln{\left\{\frac{1}{2\left(\frac{2}{\gamma(N) + \sqrt{\gamma^2(N) + 4}}\right)^{1/\beta} - 1}\right\}}\right]^{-1} + 
 H^4\left(\frac{\gamma'(N)}{8\pi^2\gamma(N)}\right)\times\nonumber\\
 &\Bigg[&\frac{\left\{1 + \frac{1}{\epsilon}\tanh\left(\frac{\epsilon\pi\alpha}{\beta GH^2}\right)\right\}^{\beta} - 
\left\{1 + \frac{1}{\epsilon}\tanh\left(\frac{\epsilon\pi\alpha}{\beta GH^2}\right)\right\}^{-\beta}}
{\alpha~\mathrm{sech}^2\left(\frac{\epsilon \pi\alpha}{\beta GH^2}\right)\left[
\left\{1 + \frac{1}{\epsilon}\tanh\left(\frac{\epsilon \pi\alpha}{\beta GH^2}\right)\right\}^{\beta-1} 
+ \left\{1 + \frac{1}{\epsilon}\tanh\left(\frac{\epsilon \pi\alpha}{\beta GH^2}\right)\right\}^{-\beta-1}\right]}\Bigg]\nonumber\\
\nonumber
 \end{eqnarray}
 and
 \begin{eqnarray}
 F_2\left[\gamma(N),\gamma'(N)\right] = H^4\left(\frac{\gamma'(N)}{8\pi^2\gamma(N)}\right)
 \left[\frac{\left\{1 + \frac{1}{\epsilon}\tanh\left(\frac{\epsilon\pi\alpha}{\beta GH^2}\right)\right\}^{\beta} - 
\left\{1 + \frac{1}{\epsilon}\tanh\left(\frac{\epsilon\pi\alpha}{\beta GH^2}\right)\right\}^{-\beta}}
{\alpha~\mathrm{sech}^2\left(\frac{\epsilon \pi\alpha}{\beta GH^2}\right)\left[
\left\{1 + \frac{1}{\epsilon}\tanh\left(\frac{\epsilon \pi\alpha}{\beta GH^2}\right)\right\}^{\beta-1} 
+ \left\{1 + \frac{1}{\epsilon}\tanh\left(\frac{\epsilon \pi\alpha}{\beta GH^2}\right)\right\}^{-\beta-1}\right]}\right]\nonumber
\end{eqnarray}
respectively. Based on Eq.(\ref{equiv potential}) and Eq.(\ref{equiv potential and LM}), 
we may argue that the entropic cosmology of $S_\mathrm{ns}$ can be equivalently 
represented by Gauss-Bonnet cosmology. 

As mentioned earlier that we consider the quasi-matter bounce scenario described by the scale factor (\ref{matter bounce-1}) to analyze the perturbation, where 
the perturbation modes generate during the contracting phase deep in the sub-Hubble regime, which in turn ensures the resolution of 
the horizon problem. The important quantities that we will need are,
\begin{eqnarray}
Q_a&=&-8\dot{h}H^2 = -\frac{4n^2(1+2n)\sqrt{\widetilde{R}}}{\pi G}\left(\frac{\widetilde{R}}{R_0}\right)^{\frac{1}{2}-n}~~~~~,~~~~~~
Q_b=-16\dot{h}H = \frac{4n(1+2n)}{\pi G}\left(\frac{\widetilde{R}}{R_0}\right)^{\frac{1}{2}-n}\, ,\nonumber\\
Q_c&=&Q_d = 0~~~~,~~~~~
Q_e=-32\dot{h}\dot{H} = \frac{8n(1+2n)\sqrt{\widetilde{R}}}{\pi G}\left(\frac{\widetilde{R}}{R_0}\right)^{\frac{1}{2}-n}~~~~,~~~~
Q_f=16 \left[ \ddot{h} - \dot{h}H \right] = 0 \, ,
\label{Q-s}
\end{eqnarray}
respectively, where $R_0 = \frac{1}{t_0^2}$ and $\widetilde{R}(t) = \frac{R(t)}{12n(1-4n)}$. 
In regard to curvature perturbation, the Mukhanov-Sasaki (MS) equation in Fourier mode comes as,
\begin{eqnarray}
\frac{d^2v_k(\eta)}{d\eta^2} + \left(k^2 - \frac{\sigma}{\eta^2}\right)v_k(\eta) = 0\, ,
\label{scalar-MS-equation}
\end{eqnarray}
here $\eta$ symbolizes the conformal time coordinate and $v(k,\eta)$ is the scalar MS variable. 
Moreover $\sigma$ is given by,
\begin{eqnarray}
\sigma = \xi(\xi - 1)\left[1 + 24\left(1-4n^2\right)\left(\frac{\widetilde{R}}{R_0}\right)^{\frac{1}{2} - n}\right]\, ,
\label{sigma}
\end{eqnarray}
which is approximately a constant during the generation era of the perturbation modes in the sub-Hubble regime during the contracting phase, 
due to the condition $n < 1/2$ (required to solve the horizon problem). In effect of which and considering 
the Bunch-Davies initial condition, 
the scalar power spectrum $\mathcal{P}_{\Psi}(k,\eta)$ in the super-horizon scale becomes,
\begin{eqnarray}
\mathcal{P}_{\Psi}(k,\eta) = \left[\left(\frac{1}{2\pi}\right)\frac{1}{z\left|\eta\right|}\frac{\Gamma(\nu)}{\Gamma(3/2)}\right]^2
\left(\frac{k|\eta|}{2}\right)^{3-2\nu}\, ,
\label{scalar-power-spectrum-superhorizon}
\end{eqnarray}

In regard to the tensor perturbation, the Mukhanov-Sasaki equation takes the following form,
\begin{align}
\frac{d^2v_T(k,\eta)}{d\eta^2} + \left(k^2 - \frac{\sigma_T}{\eta^2}\right)v_T(k,\eta) = 0\, ,
\label{tensor-MS-equation}
\end{align}
where $v_T(k,\eta)$ being the Fourier mode for the tensor MS variable, and $\sigma_T$ has the following form, 
\begin{eqnarray}
\sigma_T = \xi(\xi - 1)\left[1 - 16(1-4n^2)\left(\frac{\widetilde{R}}{R_0}\right)^{\frac{1}{2} - n}\right]\, .
\label{sigma-T}
\end{eqnarray}
Due to $n < 1/2$, the quantity $\sigma_T$ can be safely considered to be a constant during the generation era 
of the perturbation modes at the contracting phase of the universe. 
Here it may be mentioned that both the tensor polarization modes ($+$ and $\times$ polarization modes) obey the same evolution Eq.(\ref{tensor-MS-equation}) -- 
this means that the two polarization modes equally contribute to the energy density of the tensor perturbation variable, and thus 
we will multiply by the factor '2' in the final expression of the tensor power spectrum. 
Similar to the curvature perturbation variable, the tensor perturbation initiates from the Bunch-Davies vacuum at the distant past, 
i.e. $v_T(k,\eta)$, i.e $\lim_{k|\eta| \gg 1}v_T(k,\eta) = \frac{1}{\sqrt{2k}}e^{-ik\eta}$. With such initial condition, we obtain the tensor 
power spectrum for $k$th mode in the super-Hubble regime as,
\begin{align}
\mathcal{P}_{T}(k,\tau) = 2\left[\frac{1}{2\pi}\frac{1}{z_T\left|\eta\right|}\frac{\Gamma(\theta)}{\Gamma(3/2)}\right]^2 \left(\frac{k|\eta|}{2}
\right)^{3 - 2\theta}\, ,
\label{tensor-power-spectrum}
\end{align}
where $\theta = \sqrt{\sigma_T + \frac{1}{4}}$. 
Having obtained the scalar and tensor power spectra, we determine $n_s$ and $r$, and they are given by (the suffix 'h' with a quantity 
represents the quantity at the instant of horizon crossing), 
\begin{eqnarray}
n_s = 4 - \sqrt{1 + 4\sigma_h} \, , \quad r = 2\left[\frac{z(\eta_h)}{z_T(\eta_h)}\frac{\Gamma(\theta)}{\Gamma(\nu)}\right]^2
\left( k\left|\eta_h\right| \right)^{2(\nu-\theta)}\, ,
\label{obs-2}
\end{eqnarray}
where the quantities have the following forms,
\begin{align}
\nu=&\sqrt{\sigma_h + \frac{1}{4}}\, ; \quad \sigma_h = \xi(\xi - 1)\left[1 + 24\left(1-4n^2\right)\left(\frac{\widetilde{R}_h}{R_0}\right)^{\frac{1}{2} - n}
\right]\, ,\nonumber\\
\theta=&\sqrt{\sigma_{T,h} + \frac{1}{4}} \, ; \quad \sigma_{T,h} = \xi(\xi - 1)\left[1 - 16(1-4n^2)\left(\frac{\widetilde{R}_h}{R_0}\right)^{\frac{1}{2} - n}
\right]\, ,\nonumber\\
z(\eta_h)=&-\frac{1}{\sqrt{n}}\left(\frac{a_0^n}{\kappa\widetilde{R}_h^{n}}\right)\left[1 - 24n(1+2n)
\left(\frac{\widetilde{R}_h}{R_0}\right)^{\frac{1}{2} - n}\right]\, ,\nonumber\\
z_T(\eta_h)=&\frac{1}{\sqrt{2}}\left(\frac{a_0^n}{\kappa\widetilde{R}_h^{n}}\right)\left[1 + 16n(1+2n)
\left(\frac{\widetilde{R}_h}{R_0}\right)^{\frac{1}{2} - n}\right]\, .
\label{obs-3}
\end{align}
\begin{figure}[!h]
\begin{center}
\centering
\includegraphics[width=3.0in,height=3.0in]{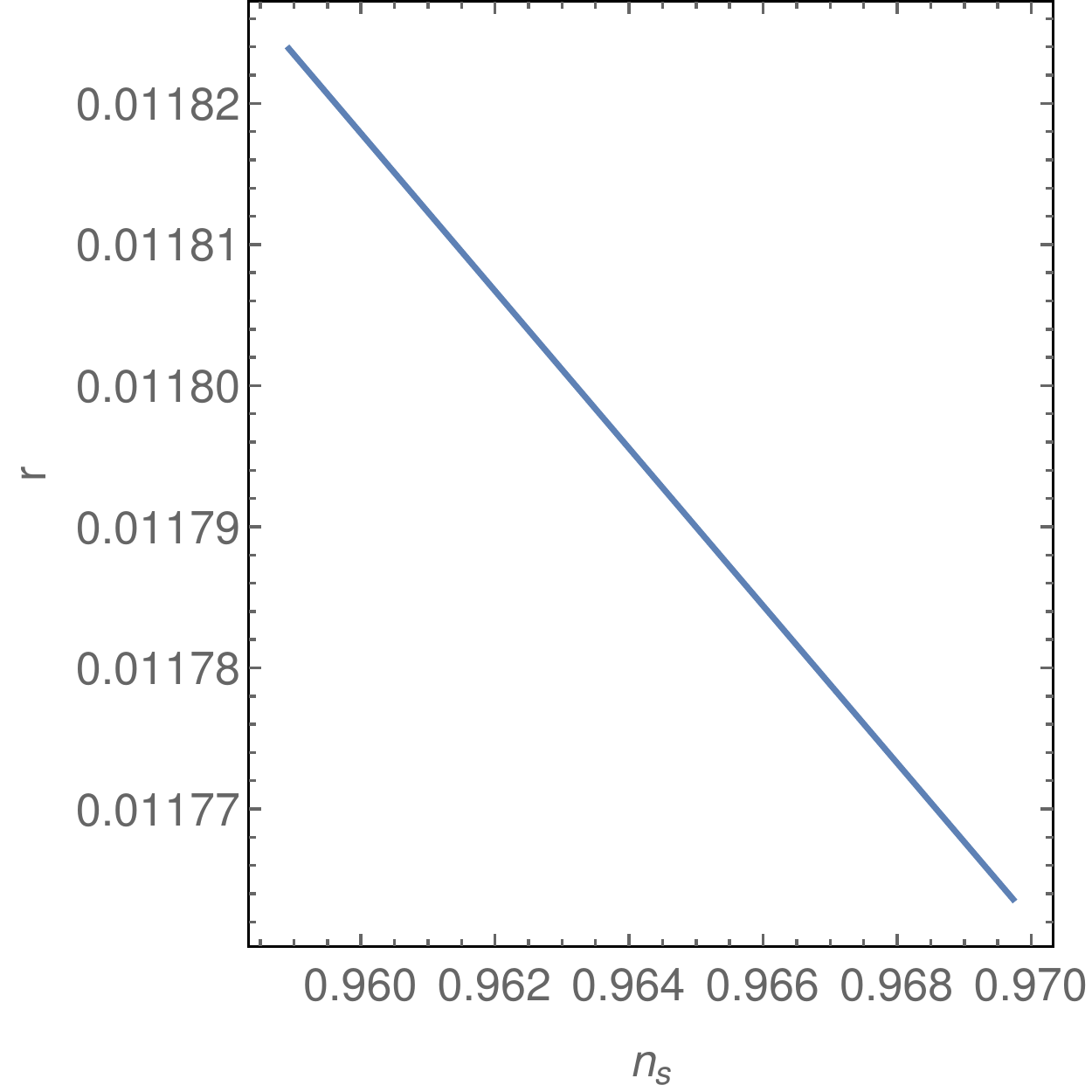}
\caption{Parametric plot of $n_s$ (along $x$-axis) vs. $r$ (along $y$-axis) 
with respect to $n$. Here we take $\alpha = [0.0938,0.0939]$ and $\beta = \frac{\pi}{16}$.}
\label{plot-observable}
\end{center}
\end{figure}

Here $\widetilde{R}_h$ represents the Ricci scalar at the horizon crossing, and using the horizon crossing condition $k\eta_h = \frac{2n}{1-2n}$, it comes 
as,
\begin{eqnarray}
\widetilde{R}_h = \left[\frac{1}{26na_0^n}\right]^{2/(1-2n)}\mathrm{By}^{-2}\, .
\label{obs-6}
\end{eqnarray}
Therefore it is clear that $n_s$ and $r$ in the present context depends on the parameters $n$ and $a_0$. Here we need to recall that $n$ 
and $a_0$ are related to the entropic parameters as $n = \sqrt{\alpha}$ and $a_0 = \pi/\left(4\beta\right)$ respectively. It turns out that the theoretical 
predictions for $n_s$ and $r$ get simultaneously compatible with the recent Planck data for a small range of the entropic parameters 
given by: $\alpha = [0.0938,0.0939]$ and $\beta = \frac{\pi}{16}$, see Fig.[\ref{plot-observable}].

\section{Conclusion}

In this short review article, we have proposed generalized entropic function(s) and have addressed their implications on 
black hole thermodynamics as well as on cosmology. In the first half of the paper, a 4-parameter and a 3-parameter generalized entropy 
functions are shown, which are able to generalize the known entropies proposed so far, like the Tsallis, R\'{e}nyi, Barrow, Sharma-Mittal, Kaniadakis and 
Loop Quantum Gravity entropies for suitable choices of the respective entropic parameters. However the 4-parameter entropy functions proves to be 
more general compared to the 3-parameter entropy function, in particular, the 3-parameter entropy does not converge to the Kaniadakis entropy for any
choices of the parameters, unlike to the entropy having 4 parameters which generalizes all the known entropies including the Kaniadakis one. Thus 
regarding to the number of parameters in a generalized entropy function, we have provided a conjecture -- 
``The minimum number of parameters required in a generalized entropy function that can generalize 
all the known entropies mentioned above is equal to four''. Consequently the interesting implications of 3-parameter entropy on black hole thermodynamics 
and the 4-parameter entropy on cosmology have been addressed. It turns out that the entropic cosmology corresponding to the 4-parameter 
generalized entropy results to an unified cosmological scenario of early inflation and the late dark energy era of the universe, where the observable 
quantities are found to be compatible with the recent Planck data for certain viable ranges of the entropic parameters.

Despite these successes, here it deserves mentioning that the 4-parameter entropy function ($S_\mathrm{g}$) 
seems to be plagued with singularity for certain cosmological evolution of the universe. In particular, $S_\mathrm{g}$ diverges 
at the instant when the Hubble parameter vanishes, for instance at the instant of bounce in the context of bounce cosmology. 
With this spirit, we have proposed a singular-free 5-parameter entropy function ($S_\mathrm{ns}$) which converges 
to all the known entropy functions for particular limits of the entropic parameters, and at the same time, 
also proves to be non-singular for the entire cosmological evolution of the universe even at $H = 0$ (where $H$ represents the Hubble parameter). 
Regarding to the non-singular entropy, a second conjecture has been given : 
``The minimum number of parameters required in a generalized entropy function that can generalize 
all the known entropies, and at the same time, is also singular-free during the universe's evolution -- 
is equal to five''. Such non-singular behaviour of $S_\mathrm{ns}$ proves to be useful in describing the bounce cosmology, in particular, the 
entropic cosmology corresponding to $S_\mathrm{ns}$ naturally allows symmetric bounce universe. With the perturbation analysis in the context of entropic 
bounce, it has been shown that the observable quantities like the spectral tilt and the tensor-to-scalar ratio are simultaneously compatible with 
the Planck data in the background of symmetric quasi-matter bounce scenario. 

Finally we would like to mention that the proposals of generalized entropy functions ($S_\mathrm{g}$ or $S_\mathrm{ns}$) opens a new directions in 
theoretical physics, and its vast consequences may hint some unexplored directions of black hole thermodynamics as well as of cosmology. 
For example, it will be of utmost interest to study the aspects of the generalized entropy functions 
on primordial black hole formation or primordial gravitational wave or the recently found astrophysical black holes as well. 
With the recent and future advancements of different detectors (like the GW detectors or regarding the black hole detection), we hope that 
these study can indirectly quantify the viable ranges of entropic parameters.

\section*{Acknowledgments}

This work was supported by MINECO (Spain), project PID2019-104397GB-I00 and  also partially supported by the program Unidad de 
Excelencia Maria de Maeztu CEX2020-001058-M, Spain (SDO). 
This research was also supported in part by the 
International Centre for Theoretical Sciences (ICTS) for the online program - Physics of the Early Universe (code: ICTS/peu2022/1) (TP).

\end{document}